\documentstyle[prl,aps,multicol,epsf]{revtex}
\begin{document}
\draft
\title{Conductivity of Silicon Inversion Layers: comparison with and without
in-plane magnetic field.}

\author{Yeekin Tsui, S. A. Vitkalov and M. P. Sarachik}
\address{Physics Department, City College of the City
University of New York, New York, New York 10031}
\author{T.~M.~Klapwijk}
\address{Kavli Institute of Nanoscience, Delft University of Technology,
Department of Applied Physics, 2628 CJ Delft, The Netherlands}
\date{\today}

\maketitle

\begin{abstract}

A detailed comparison is presented of the temperature dependence of the
conductivity of dilute, strongly interacting electrons in two-dimensional
silicon inversion layers in the metallic regime in the presence and in
the absence of a magnetic field.  We show explicitly and quantitatively that a
magnetic field applied parallel to the plane of the electrons reduces the
slope of the conductivity versus temperature curves to near zero over a broad
range of electron densities extending from $n_c$ to deep in the metallic
regime where the high field conductivity is on the order of $10 e^2/h$.  The
strong suppression (or "quenching") of the metallic behavior by a magnetic
field sets an important constraint on theory.

\end{abstract}


\begin{multicols}{2}

The conductivity of low density, strongly interacting electrons (or holes) in
two dimensions increases with decreasing temperature above a critical electron
density $n_c$ (or hole density $p_c$), raising the possibility that
there exists an unexpected metallic phase and a metal-insulator transition in
two dimensions \cite{review}.  This behavior has been observed in many
different 2D systems and is particularly pronounced in inversion layers in
silicon MOSFET's.  The application of a magnetic field parallel to the plane
of the electrons (or holes) has a dramatic effect, causing the conductivity
to change by many orders of magnitude at low temperatures and low densities
near $n_c$ ($p_c$).  In silicon MOSFET's, the conductivity decreases as the
magnetic field is increased and then saturates to a value that is
approximately constant \cite{Simonian97,Pudalov97}.  Other systems exhibit
very similar behavior, with a conductivity that reaches a knee and then
continues to decrease but with much smaller slope \cite{Yoon99}. 
Shubnikov-deHaas experiments have been performed that indicate that the
electrons become fully polarized at or near the value of in-plane
magnetic field that causes the saturation or knee observed in the
magnetoconductivity \cite{okamoto,vitkalovSdH,Tutuc01}.  These intriguing and
quite anomalous effects have been the subject of a great deal of interest and
debate.

Although a number of studies have shown qualitatively that a magnetic
field decreases the conductivity and suppresses the metallic behavior
\cite{Simonian97,Dolgopolov92,Mertes01,Shashkin01}, there has been no
systematic investigation of the temperature dependence in moderate and
high magnetic field.  The purpose of the present note is to demonstrate
explicitly and quantitatively that the application of a magnetic field
parallel to the plane of the electrons in silicon inversion layers sharply
reduces the temperature dependence of the conductivity over a broad range
extending to electron densities deep in the metallic regime where the
conductivity at high field is on the order of $10e^2/h$.

Data are presented for three silicon MOSFETs with mobilities $\mu$ at
$4.2$ K of $\approx 30,000\;$V/(cm$^2s)$ (sample \#$1$) and
$20,000\;$V/(cm$^2s)$ (samples \#$2$ and \#$3$).  Contact resistances were
minimized by using a split-gate geometry, which allows a higher electron
density in the vicinity of the contacts than in the 2D system under
investigation.  The resistance was measured in a $^3$He Oxford Heliox
system as a function of temperature in zero field and in a parallel field
of $10$ Tesla by standard four-probe $AC$ techniques using currents in
the linear regime, typically below $5 nA$, at frequency 3Hz.  Metallic
temperature dependence was found in zero field for all samples at electron
densities above $n_c \approx 0.9 \times 10^{11}$ cm$^{-2}$.

The conductivity of a silicon MOSFET sample in the absence of magnetic field
is shown as a function of temperature for eight different electron
densities in Fig. 1(A); Fig. 1(B) shows the conductivity for the same electron
densities in a magnetic field of $10$ Tesla applied parallel to the electron
plane.  The temperature dependence in the absence of a field is strongly
suppressed by an in-plane magnetic field of 10 T.  Similar results  were
obtained for the two other samples studied.  It should be noted that the
conductivity is near or at its high-field, saturated value in 10 T for all
the densities shown.

In order to demonstrate the effect of high in-plane magnetic fields, we
determined the slope of the conductivity curves $d\sigma/dT$ in zero field and
in high magnetic field; this is illustrated in Fig. 2 for two different
electron densities.

We now examine how the temperature dependence of the conductivity evolves as
the magnetic field is increased from zero to a value high enough that the
conductivity has reached its high-field saturated value where the electron
spins are completely aligned.  For a constant electron density $n = 1.64
\times 10^{11}$ cm$^{-2}$, the closed symbols of Fig. 3 are the slopes
$d\sigma(H)/dT$ of the
$\sigma$ versus $T$ curves for several values of in-plane magnetic field. 
Although the $\sigma$ versus
$T$ curves at finite field exhibit some detailed structure \cite{structure},
these are small effects that do not affect the over-all behavior in a
substantial way.  For comparison, the open
symbols in Fig. 3 denote the difference $[\sigma(1.35 K) -\sigma(0.27 K)]$;
the main features of the curve are unaltered.  The (negative) slope changes
rapidly with increasing in-plane magnetic field and assymptotically
approaches a value near zero as the field approaches the value required to
saturate the conductivity and align the spins.

\vbox{
\vspace{0.4 in}
\hbox{
\hspace{-0.2in} 
\epsfxsize 3.4 in \epsfbox{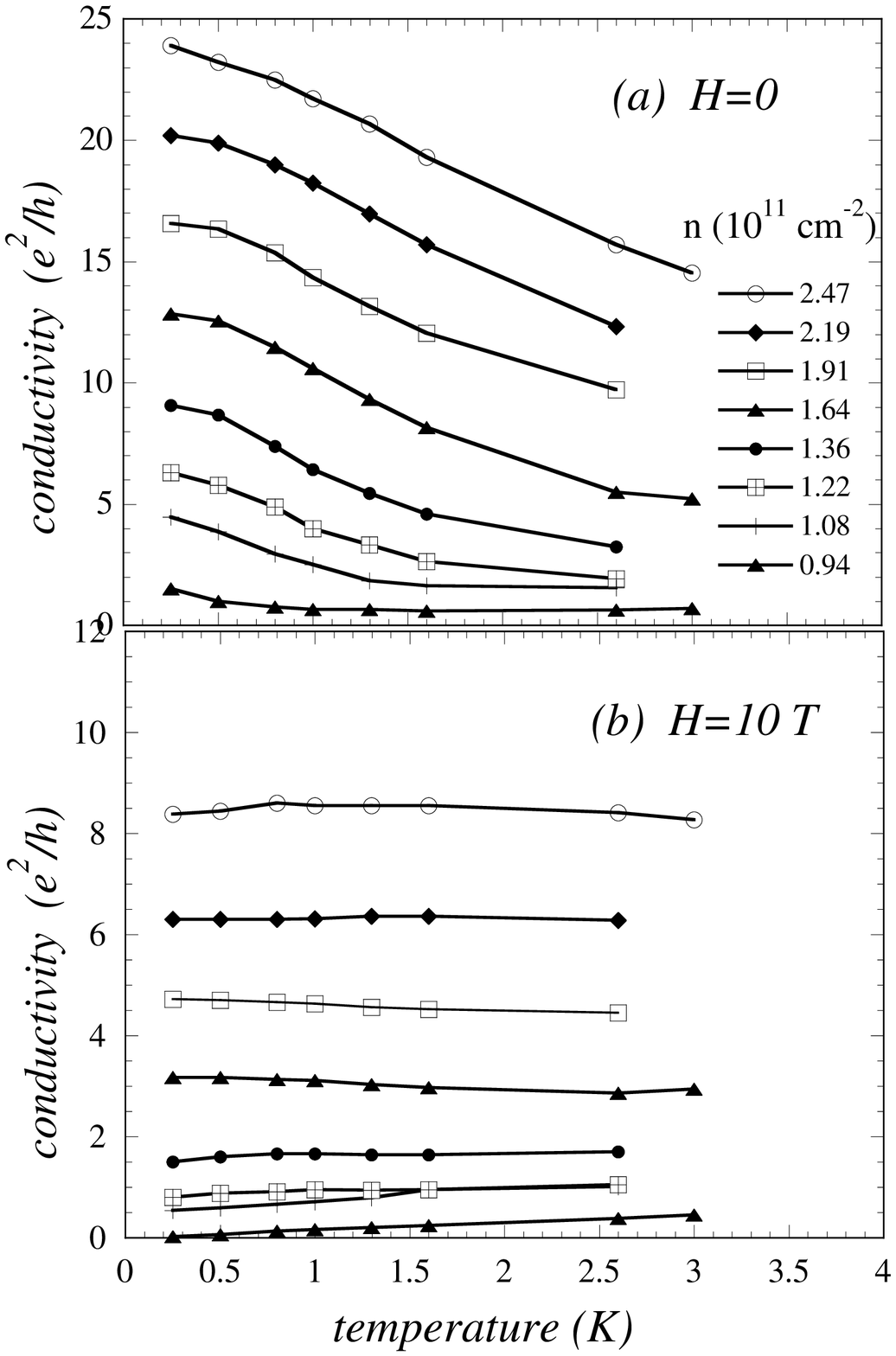} 
}
}
\vskip 0.5cm
\refstepcounter{figure}
\parbox[b]{3.1in}{\baselineskip=12pt FIG.~\thefigure.
Conductivity of silicon MOSFET sample \#$2$ as a function of temperature
for different electron densities, as labelled: (A) in the absence of
external magnetic field; (B) in a field of 10 T applied parallel to the
plane.  Similar results were obtained for samples \#$1$ and \#$3$.

\vspace{0.10in}
}
\label{raw}

The behavior illustrated in Fig. 3 obtains over a broad range of electron
densities deep into the metallic phase, where the conductivity is $10$ to
$20$ times the quantum unit of conductance.  Figure 4 shows the ratio of
the slope in a high in-plane field of $10$ T to the slope in zero
field, $\frac{d\sigma(H=10 T)/dT}{d\sigma(0)/dT}$, plotted as a function
electron density.  As indicated by the dashed horizontal lines, the ratio
does not exceed $\pm 0.1$ and is near zero over the entire range of densities
studied, from $n = 0.95 \times 10^{11}$ cm$^{-2}$ to $2.5 \times 10^{11}$
cm$^{-2}$.   The temperature dependence in high in-plane magnetic field
exhibits some scatter.  It is weak but finite, and it is positive or negative
depending on electron density.  We attribute this to remanent, weak effects
that become dominant when the field has suppressed the strong zero-field
dependence on temperature.

\vbox{
\vspace{0.4 in}
\hbox{
\hspace{-0.2in} 
\epsfxsize 3.4 in \epsfbox{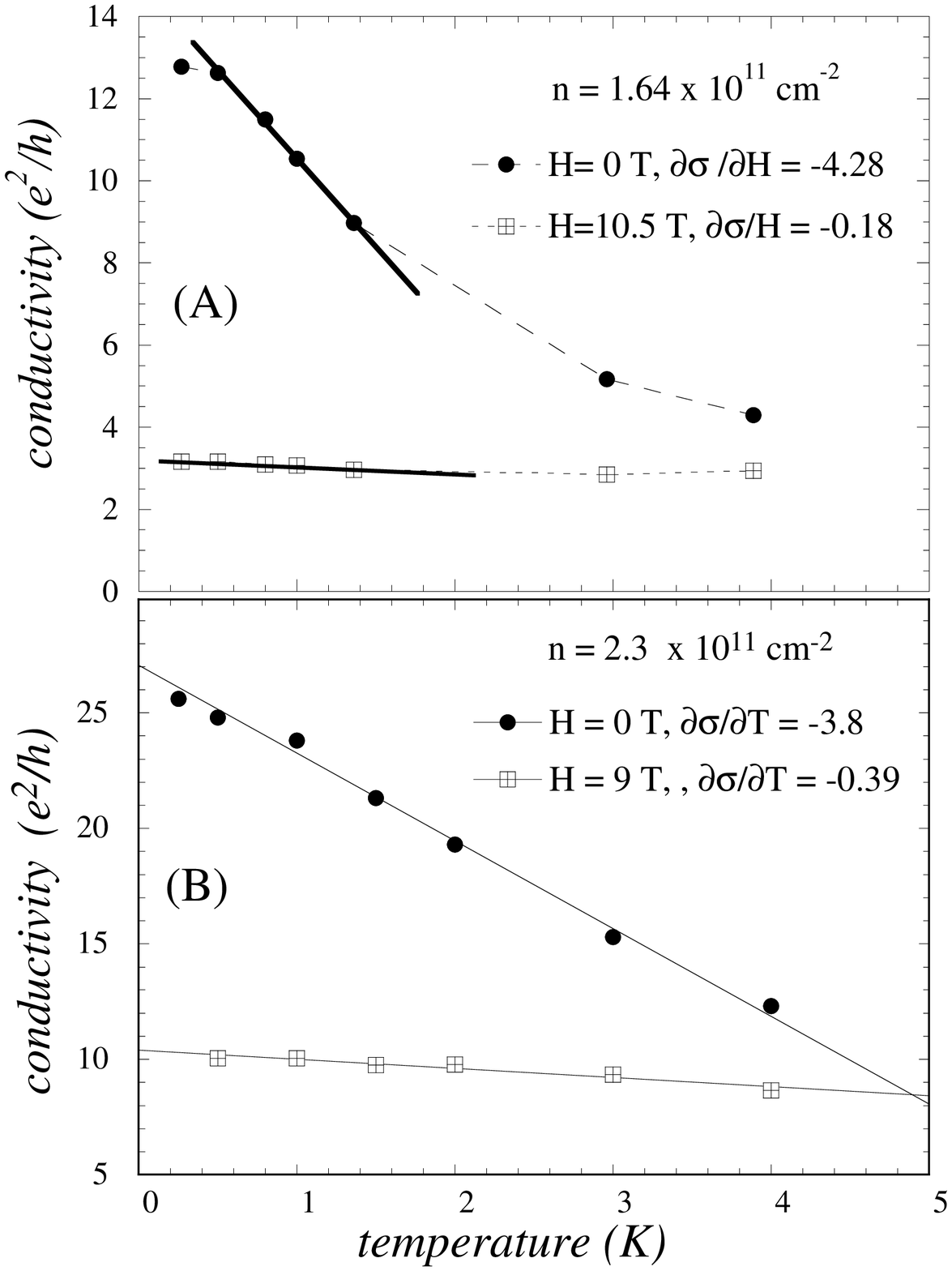} 
}
}
\vskip 0.5cm
\refstepcounter{figure}
\parbox[b]{3.1in}{\baselineskip=12pt FIG.~\thefigure.
For two electron densities, the lines illustrate the procedure used to
determine the slope $d\sigma/dT$ plotted in Fig. 4.  Note that the range
over which the slope is approximately constant broadens as the electron
density is increased.  Data shown for sample \#$3$.

\vspace{0.10in}
}
\label{metallic}

Many theories have been proposed to account for the interesting behavior of
two-dimensional systems of electrons such as silicon MOSFET's. 
Temperature-dependent screening in a Fermi gas has been suggested by many
\cite{screening} as the source of the temperature dependence of the
conductivity.  A recent theory of Zala, Narozhny and Aleiner
\cite{Zala}, which considers exchange as well as Hartree terms, provides
sufficient detail to allow \vbox{
\vspace{0.4 in}
\hbox{
\hspace{-0.2in} 
\epsfxsize 3.4 in \epsfbox{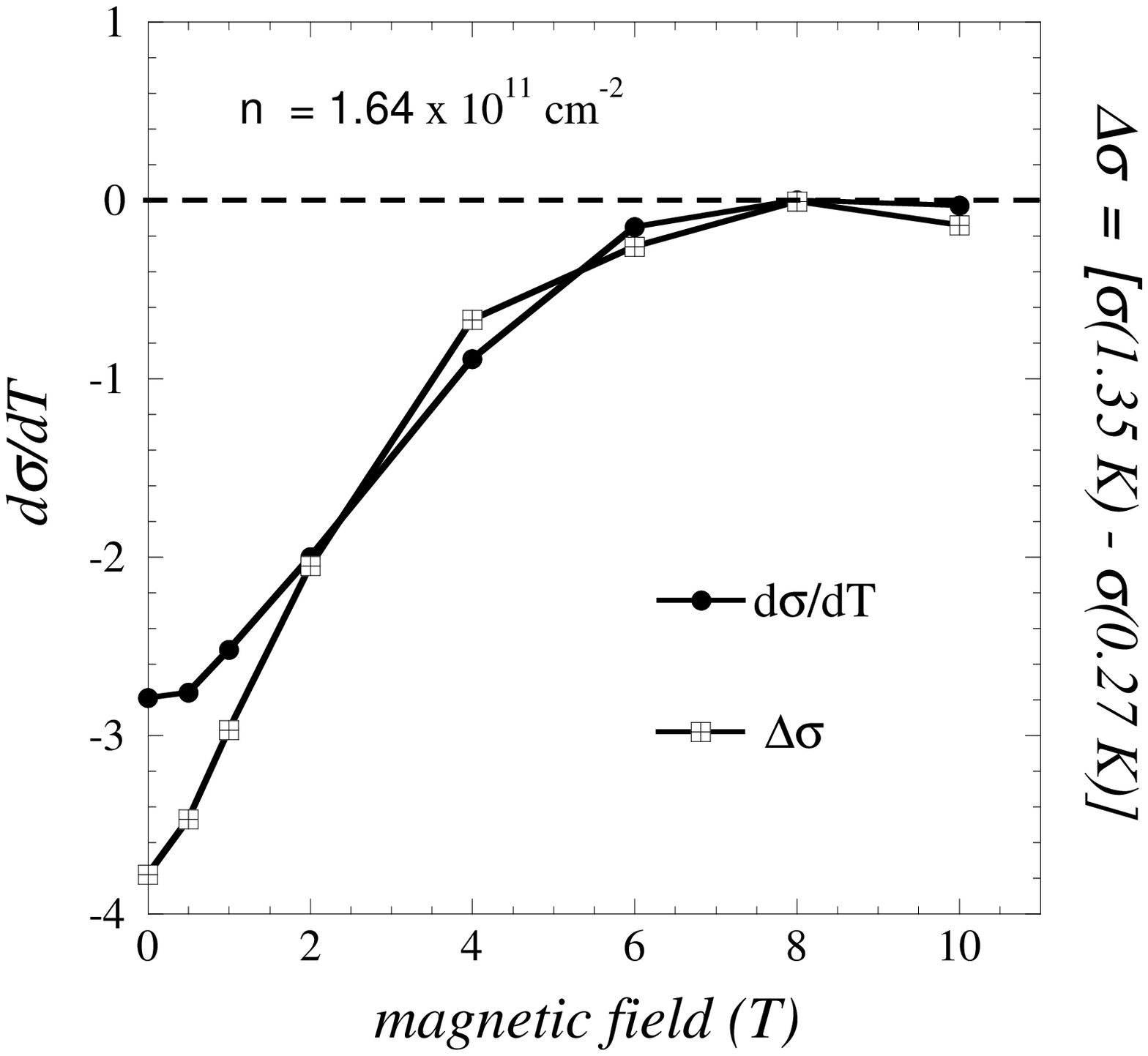} 
}
}
\vskip 0.5cm
\refstepcounter{figure}
\parbox[b]{3.1in}{\baselineskip=12pt FIG.~\thefigure.
Closed symbols denote the slope $d\sigma/dT$ versus in-plane magnetic field
$H$ for silicon MOSFET sample \#$2$ at electron density $1.64 \times 10^{11}$
cm$^{-2}$.  Open symbols denote the difference in conductivity [$\sigma(1.37
K) - \sigma(0.27 K)$].

\vspace{0.30in}
}
\label{delta} comparison with experiment.  Within this theory,
the ratio plotted in Fig. 4 can be
expressed in term of the Fermi liquid parameter $F_0^\sigma$:

$$
r = \frac{d\sigma(H)/dT}{d\sigma(0)/dT} = 
\frac{1+[3F_0^\sigma/(1+F_0^\sigma)]}{1+[15F_0^\sigma/(1+F_0^\sigma)]}.
$$

The inset to Fig. 4 shows a plot of the ratio $r =
[(d\sigma(H)/dT)/(d\sigma(0)/dT)]$ plotted as a function of the Fermi liquid
parameter $F_0^\sigma$.

Direct comparisons of the temperature and field dependence of the conductivity
with the theory of Zala {\it et al.} have yielded a range of values for the
Fermi liquid parameter $F_0^\sigma$.  Vitkalov $et al.$
\cite{Vitkalov} obtained $F_0^\sigma= -0.15$ for electron densities between
$1.6 \times 10^{11}$ cm$^{-2}$ and $3.3 \times 10^{11}$ cm$^{-2}$. 
Examination of the inset to Fig. 4 shows that this is inconsistent with the
ratios between $0.1$ and $-0.1$ found experimentally.  A value near
$F_0^\sigma= -0.5$ obtained in this range of densities by Pudalov $et al.$
\cite{Pudalov} is marginally outside the bounds shown in Fig. 4, while
$F_0^\sigma \approx -0.3$ reported by Shashkin {\it et al.} \cite
{Shashkin02} falls between the required bounds.

A number of other theoretical scenarios have been advanced, including
percolation in an inhomogeneous system composed of metallic and insulating
regions, a Wigner crystal or glass, ferromagnetism, superconductivity,
a spin glass, an electron glass \cite{review}.  Complete suppression of
the zero-field temperature dependence by in-plane magnetic field is
predicted by the theory of Spivak and Kivelson \cite{spivak}, which considers 
phase separation and intermediate phases between the Fermi liquid and the
Wigner crystal.

\vbox{
\vspace{0.4 in}
\hbox{
\hspace{-0.2in} 
\epsfxsize 3.4 in \epsfbox{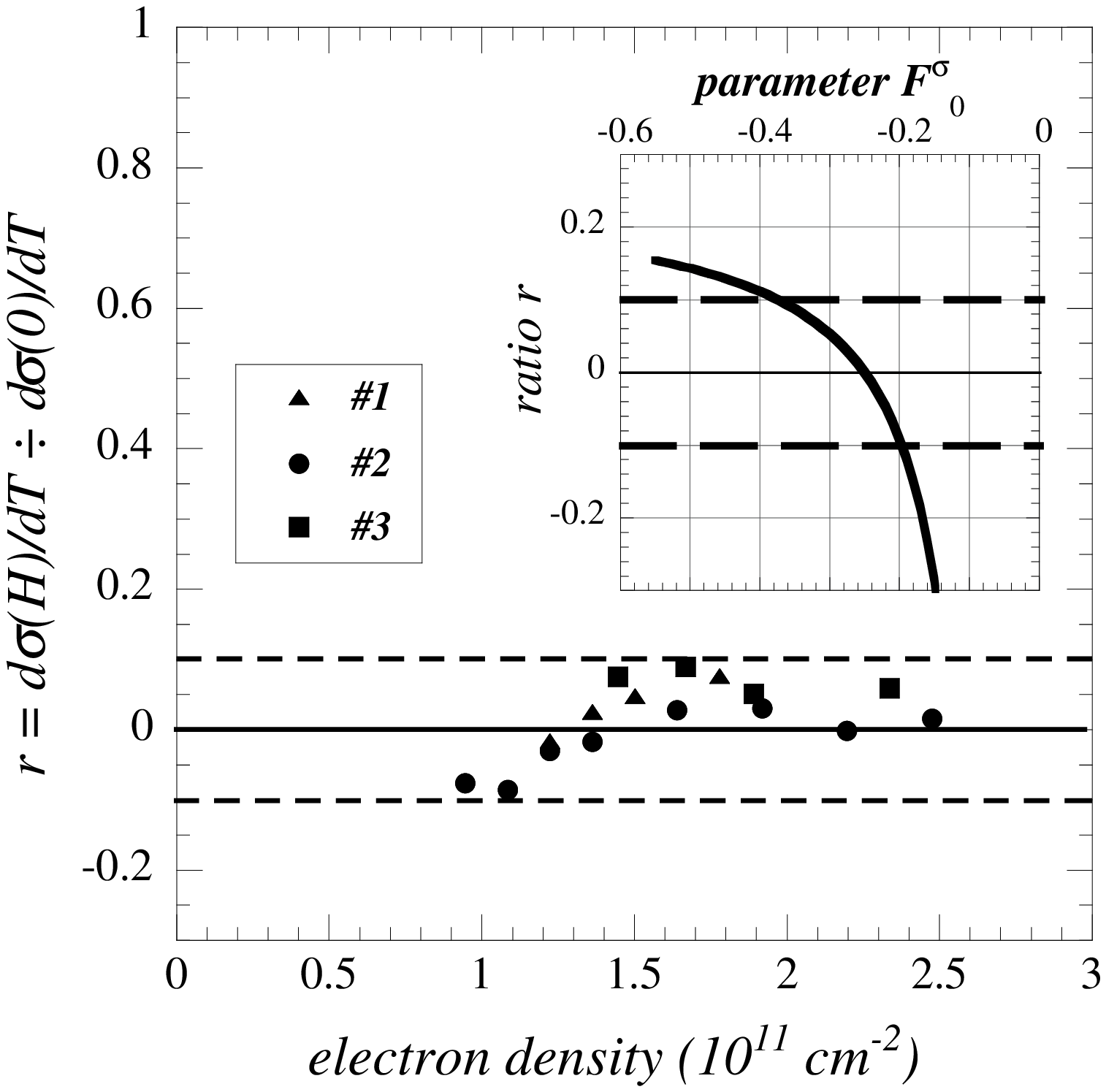} 
}
}
\vskip 0.5cm
\refstepcounter{figure}
\parbox[b]{3.1in}{\baselineskip=12pt FIG.~\thefigure.
The ratio $r = \frac{d\sigma(H)/dT}{d\sigma(0)/dT}$ versus electron density
for three samples, as labelled.  The inset shows $r$ versus the Fermi liquid
parameter $F_0^\sigma$ predicted by the theory of Zala {\it et al.}
\cite{Zala}.

\vspace{0.10in}
}
\label{delta}

In summary, data are reported for inversion layers in high mobility
silicon MOSFET's that demonstrate that an in-plane magnetic field
suppresses the metallic temperature dependence of the conductivity observed 
in the absence of magnetic field.  The metallic behavior
is strongly suppressed (or "quenched") over a broad range of
densities extending from $n_c$ upward deep into the metallic regime where the
high field conductivity is $10$ times the quantum unit of
conductivity.  This is a robust and central feature of these two-dimensional
systems that sets an important constraint on theory.

We thank Boris Spivak and Steve Kivelson for stimulating this paper and for
numerous discussions.  This work was supported by DOE-FG02-84-ER45153 and NSF
grant DMR-0129581.

\end{multicols}

\end{document}